%% file: ms.tex
\documentclass[conference,a4paper, final]{IEEEtran}
\pdfoutput=1
\usepackage[nolist]{acronym}
\usepackage{tikz}
\usetikzlibrary{decorations.pathreplacing,calligraphy}
\usepackage{pgfplots}
\usepgfplotslibrary{groupplots}
\usepackage{graphicx}
\usepackage{subfloat}
\usepackage[font=small]{caption}
\usepackage{mathtools}
\usepackage{subcaption}
\usepackage{float}
\pgfplotsset{compat=newest}
\usepackage{units}
\usepackage{amsmath, amsbsy, amssymb}
\usepackage{tikzscale}
\usepackage{color}
\usepackage{epigraph}
\usepackage{circuitikz}
\usepackage{multirow}
\usepackage{booktabs}
\usepackage[ruled]{algorithm2e}
\usepackage{adjustbox}
\usepackage[group-separator={,}]{siunitx}
\definecolor{darkgreen}{rgb}{0.125,0.5,0.169}
\usetikzlibrary{shapes,arrows,fadings,chains}
\usepackage{marvosym}
\usepackage{bbm}
\usepackage{mathtools}
\usetikzlibrary{positioning,fit,calc}
\usetikzlibrary{calc,patterns,angles,quotes}
\usetikzlibrary{shapes.geometric, arrows}
\usetikzlibrary{fit}
\usetikzlibrary{shapes.multipart}
\usetikzlibrary{calc, positioning}
\usetikzlibrary{decorations}
\usetikzlibrary{decorations.pathreplacing}
\usetikzlibrary{arrows.meta,shapes.arrows}
\usetikzlibrary{shapes}
\usetikzlibrary{positioning,fit,backgrounds}
\tikzset{>=latex}
\usepackage{tikzit}

\input{macros.tex}
\input{acronyms.tex}
\input{corporateColours.tex}

\input{plotcyclelists.tex}
\input{system.tikzstyles}

\IEEEoverridecommandlockouts

\begin{document}

\title{Deep Reinforcement Learning\\ for mmWave Initial Beam Alignment}

\author{\IEEEauthorblockN{Daniel Tandler, Sebastian D\"orner, Marc Gauger and Stephan ten Brink}

\IEEEauthorblockA{
 Institute of Telecommunications, University of Stuttgart, Pfaffenwaldring 47, 70659 Stuttgart, Germany \\
\{tandler,doerner,gauger,tenbrink\}@inue.uni-stuttgart.de\\
}

}

\maketitle

\begin{abstract}
We investigate the applicability of deep reinforcement learning algorithms to the adaptive initial access beam alignment problem for mmWave communications using the state-of-the-art proximal policy optimization algorithm as an example. In comparison to recent unsupervised learning based approaches developed to tackle this problem, deep reinforcement learning has the potential to address a new and wider range of applications, since, in principle, no (differentiable) model of the channel and/or the whole system is required for training, and only agent-environment interactions are necessary to learn an algorithm (be it online or using a recorded dataset). We show that, although the chosen off-the-shelf deep reinforcement learning agent fails to perform well when trained on realistic problem sizes, introducing action space shaping in the form of beamforming modules vastly improves the performance, without sacrificing much generalizability. Using this add-on, the agent is able to deliver competitive performance to various state-of-the-art methods on simulated environments, even under realistic problem sizes. This demonstrates that through well-directed modification, deep reinforcement learning  may have a chance to compete with other approaches in this area, opening up many straightforward extensions to other/similar scenarios.
\end{abstract}

\acresetall

\section{Introduction}

MmWave communication systems are usually equipped with large antenna arrays to enable highly directional transmissions using beamforming techniques to overcome the large pathloss at these frequencies. This large number of antennas paired with the usage of hybrid or analog beamforming to limit cost and power consumption raises the need of new \ac{IA} strategies for mmmWave systems, often referred to as beam training or \ac{BA}. Most of the existing work done on BA relies on using beam codebooks, i.e. employing a predefined set of beampatterns, together with methods like compressed sensing, Bayesian approaches and machine learning. Recently, it was demonstrated that not using predefined codebooks together with unsupervised deep learning may offer performance improvements compared to codebook-based approaches \cite{BayesDNN}.

There has been a great amount of work concerning the mmWave \ac{BA} problem using codebooks of predefined beampatterns and mostly classical methods, like in \cite{Codebook_2014}, where a hierarchical codebook was designed and used in conjunction with compressed sensing methods to perform \ac{CSI} estimation followed by precoder/combiner design using this estimate. In \cite{ActiveLearning} the power-based bisection search applied in \cite{Codebook_2014} is enhanced to a noisy generalized form by making use of the Bayes posterior. The work in \cite{AgileLink} introduced AgileLink, a fast mmWave \ac{BA} algorithm using hashing of random direction to determine the best beam alignment in a logarithmic number of measurements. Other approaches like \cite{Bandit1} and \cite{Bandit2} reformulate the mmWave \ac{BA} problem into a multi armed-bandit problem and solve it using methods developed for this setting. Another line of research uses \ac{DRL} to address various related problems in the area of mmWave communications, like \cite{multiUE_drl} where a special \ac{DRL} structure is proposed to perform  $\mu$BS selection and angle-based \ac{BA}. \cite{rl_beam_codebook} proposes a method to learn and adapt specialized beam codebooks using \ac{DRL} without requiring any explicit channel knowledge.
   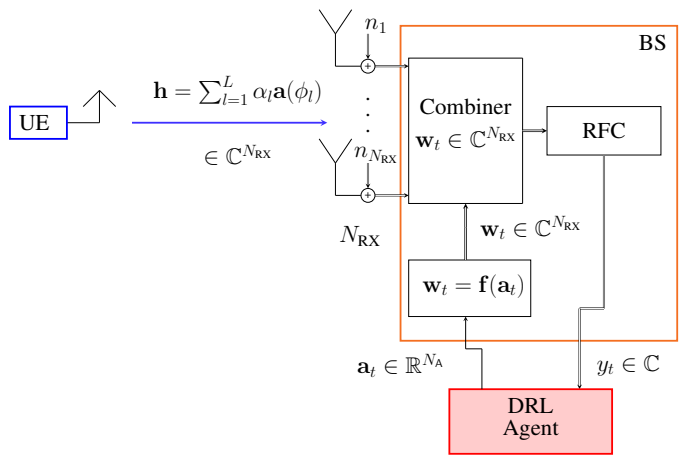
\begin{figure}[!t]
   \vspace{-0.2cm}
  \centering
    \resizebox{1\columnwidth}{!}{\input{tikz/system_model.tikz}}
 \caption{\small System model of the DRL-based BA algorithm. Note how at each timestep $t$, the DRL agent receives one complex-valued symbol $y_t$ from the receiver and emits one real valued action vector $\mathbf{a}_t$.}
 \label{fig:system_model_ba}
 \vspace{-0.5cm}
\end{figure}
The works concerning the \ac{IA} \ac{BA} problem often have in common that they choose the beams out of predefined codebooks in some way or the other. In \cite{BayesDNN} it was shown that \ac{DNN}-based codebook-free \ac{BA} algorithms may be able to outperform their codebook based counterparts. Finally, \cite{RNNE2E} shows that \acp{RNN} together with end-to-end learning are able to outperform  state-of-the-art BA methods in the codebook free (adaptive) beam alignment problem. \cite{RNNE2E} can be regarded as being the most relevant for this work as both share the same setup, optimization goals and train the same \ac{RNN}-based agent. Yet, our approach differs in the way the \ac{RNN} agent is trained: While \cite{RNNE2E} trains the agent in a end-to-end fashion requiring a specific channel model, in this work the agent is trained using the \ac{DRL} framework, where the environment is treated as a black box. This means that our \ac{DRL}-based approach could in theory also be trained directly in deployed systems with various unknown hardware impairments. Fig. \ref{fig:system_model_ba} depicts the system under consideration.
The main contributions of this work are as follows:
\begin{itemize}
    \item We demonstrate the difficulties that arise when applying off-the-shelf state-of-the-art \ac{DRL} algorithms to the problem of adaptive \ac{IA} \ac{BA} with the example of the \ac{PPO} \cite{PPO} algorithm
    \item We propose a modification of the \ac{DRL} algorithm by introducing beamforming modules to address these difficulties, which can in principle be applied to all other continuous-valued action space \ac{DRL} algorithms
    \item We show that by using this modification, the \ac{PPO} algorithm performs well even under realistic system sizes
\end{itemize}
We want to emphasize that our objective is not to outperform the model-based end-to-end trained state-of-the-art methods of \cite{BayesDNN} and \cite{RNNE2E} (which would be quite unrealistic since we essentially try to approximate the training gradient of these methods with the \ac{DRL} framework), but to show that the \ac{DRL} approach can also deliver comparably good performance \emph{without} the explicit need for any knowledge or model of the mmWave channel during training.

This paper is organized as follows: Section \ref{sec:prelim} describes the system model and gives some information about the basics of \ac{DRL}. In Section \ref{sec:drl}, we propose a way to reformulate the \ac{BA} problem into a \ac{POMDP} and discuss possible mappings of the action space of the \ac{DRL} agent to the space of combining vectors. Following this, Section \ref{sec:result} gives some numerical results and compares the performance of the \ac{DRL} approach to various baselines. Finally, Section \ref{sec:conclusion} concludes the paper.

\section{Preliminaries}
\label{sec:prelim}
In this paper, we consider the problem of one-sided beam alignment in mmWave communications, i.e., the initial (uplink) communication between a \ac{BS} equipped with an \ac{ULA} consisting of $N_{\text{RX}}$ antennas and a single \ac{UE} with a single antenna sending a constant signal omnidirectionally. Note that in this setup, no initial knowledge about the channel between \ac{BS} and \ac{UE} is assumed. In the rest of this paper, we assume that the \ac{BS} is only equipped with a single \ac{RF} chain, i.e., analog beamforming and that the transmission channel between UE and BS is static, i.e. constant during each run of the \ac{BA} algorithm. Also, the \ac{BS} has $T-1$ discrete time steps to probe the channel between itself and the \ac{UE} with analog probing beams $\mathbf{w}_t$ ($0\leq t < T$) to gain as much information about the channel as possible, after which the \ac{BS} has to use this gained information to output the estimated best combining vector maximizing the beamforming gain $\lVert \mathbf{w}
^H \mathbf{h}\rVert_2^2$. As shown in \cite{RNNE2E} this problem can be regarded as an active sensing problem in which the design of the $t-$th probing beam depends on all previous timesteps, i.e., $\mathbf{w}_t = \mathcal{G} (\mathbf{w}_{0 \leq t' \leq t-1}, y_{0 \leq t' \leq t-1})$ with received symbols $y_t$ and nonlinear map $\mathcal{G}$. The rest of this section gives a detailed description of the system model and the specific setting considered.

\subsection{System Model}
Since the mmWave channel can be regarded as being spatially sparse, we assume the widely used geometric channel model \cite{channel_model} for the mmWave channel description. As we consider one sided \ac{BA}, the channel can be described with a single vector, i.e.,
\begin{equation}
    \mathbf{h} = \sum_{l = 0 }^{L-1} \alpha_l \mathbf{a}(\phi_l) \in \mathbb{C}^{N_{\text{RX}}}
\end{equation}
for a channel with $L$ paths, complex path gain $\alpha_l \sim \mathcal{CN}(0,1)$ and \ac{AoA} $\phi_l \sim \mathcal{U}[-60^{\circ},60^{\circ}]$ of path $l$. $\mathbf{a}(\phi)$ describes the array response of the \ac{ULA} for angle $\phi$ and is given by
\begin{equation}
\mathbf{a}(\phi) = \frac{1}{\sqrt{N}}  (1, e^{j \pi \sin(\phi)}, \dots, e^{j\pi(N-1)\sin(\phi)}) \in \mathbb{C}^{N}
\label{eq:antenna_array_response}
\end{equation}
with $N$ antennas, and under the assumption of an element spacing of $\frac{\lambda}{2}$. As it is assumed that the \ac{BS} uses only one \ac{RF} chain and the \ac{UE} sends a constant stream of pilot symbols with normalized power $P_{\text{TX}} = 1$, the whole system at timestep $t$ can be described by
\begin{equation}
    y_t = \mathbf{w}_t^H \mathbf{h}+\mathbf{w}_t^H \mathbf{n}_t \, \in \mathbb{C}
\end{equation}
with $y_t$ being the complex-valued received symbol at the \ac{BS}, $\mathbf{w}_t$ being the $N_{\text{RX}}$ dimensional complex-valued analog combining vector and noise vector $\mathbf{n}_t \sim \mathcal{CN}(0,\sigma_n^2 \mathbf{I}_{N_{\text{RX}}})$, all for timestep $t$. The per-antenna \ac{SNR} is defined as $\text{SNR} = \frac{1}{\sigma_n^2}$. The whole system model is also depicted in Fig. \ref{fig:system_model_ba}.

\subsection{Deep Reinforcement Learning}
In this section, we provide a short overview over the basics of \ac{DRL}. The \ac{RL} framework consist of an abstract agent (sequentially) interacting with its environment to fulfill a certain task. Formally, this process can be modelled using a Markov decision process (MDP) \acused{MDP} $\mathsf{M}$ which can be expressed by the $4-$ tuple $(\mathsf{S},\mathsf{A},\mathsf{T},\mathsf{R})$, where $\mathsf{S}$ denotes the set of all possible environment states, $\mathsf{A}$ is the set of all possible actions the agent can take, $\mathsf{T}$ the transition probability function of the environment and $\mathsf{R}$ the reward probability function. In this formalism, at time step $t$, the agent observes the environment state $s \in \mathsf{S}$, executes the action $a \in \mathsf{A}$, after which the environment emits the scalar reward $r \sim \mathsf{R}(r|s,a)$, and transitions to a new environment state $s'\sim \mathsf{T}(s'|s,a)$ according to transition function $\mathsf{T}$, and the step counter increases by $1$. \acp{POMDP}, can be seen as a generalization of \acp{MDP} to scenarios where the agent does not receive the true environment state $s'$ after executing action $a$, but only an (possibly noisy) observation $o$ of said state, $o \sim O(o |s',a)$ with conditional observation probability $O$. Note that other quantities of \acp{POMDP} behave exactly like they do in \acp{MDP}. Due to only receiving partial information about its environment in the form of observations $o$, the agent is forced to act under uncertainty over its environment, which can be represented using so called belief states. The goal of an \ac{RL} algorithm in an \ac{MDP} is now to find a policy $\pi(a|s)$ (i.e., a potentially probabilistic mapping from states to actions) which maximizes the expected long-term cumulative discounted reward, defined as
\begin{equation}
    J(\pi) = \mathbb{E}_{s_0,\pi,T} \left[\sum_{t=0}^\infty \gamma^t r(s_t,a_t)\right]
\end{equation}
with initial state distribution $s_0 \sim p_0$, when following policy $\pi$ and with discount factor $\gamma \in [0,1]$ which determines how important short term rewards are versus long term rewards. Out of the vast family of different \ac{DRL} algorithms, the \ac{PPO} algorithm was chosen for this work, a state-of-the art model-free on-policy algorithm. It belongs to the class of policy gradient algorithms which directly try to optimize the policy by estimating the gradient of the reward with respect to the policy parameters. Since this algorithm has the on-policy property, it needs to interact continuously with its environment in order to improve its policy. The reason for choosing this algorithm lies in its demonstrated robustness against hyperparameter variations while still offering good performance. Also, as it was shown that DRL algorithms combined with a memory network in the form of an \ac{LSTM} network may be able to outperform their memory-less counterparts in \acp{POMDP} \cite{DRLPOMDP}, the chosen \ac{PPO} algorithm was modified in a similar way as described in \cite{DRLPOMDP}. For the implementation in this paper, we chose \ac{GRU} cells over  \ac{LSTM} cells for the memory network due to their lower complexity and ease of implementation. 
\section{Applying DRL to mmWave Beam Alignment}
\label{sec:drl}
\begin{figure}[!t]
    \centering
    \resizebox{1\columnwidth}{!}{\input{tikz/agent_env.tikz}}
     \caption{\small Agent-environment interaction for the partially observable Markov decision problem formulation of the beam alignment task.}
    \label{fig:agent_environment_interaction_ba}
    \vspace{-0.7cm}
\end{figure}
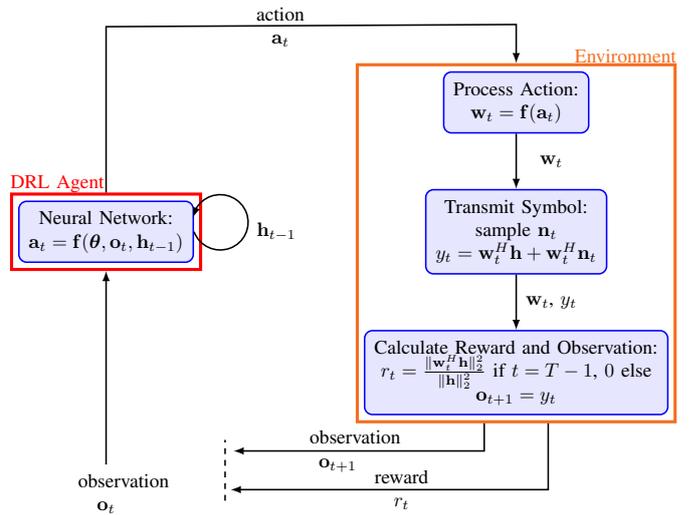

This section goes into more depth on how to combine the aforementioned \ac{RL} framework with the problem of \ac{BA}. The first task is to reformulate the \ac{BA} problem as a \ac{MDP} or \ac{POMDP} in order to enable the application of \ac{RL} algorithms. Luckily, the problem formulation offers itself very naturally to be expressed in the form of a \ac{POMDP}. While there are multiple ways to define said \ac{POMDP}, one direct way is:
\begin{itemize} \itemsep0.5em
\item $\mathsf{S}$: The set of possible channel parameters, constant over one run of the \ac{BA} algorithm
\item $\mathsf{A}$: To be specified
\item $\mathsf{T}$: $\mathsf{T}(s'|s,a) =\left\{\begin{array}{ll} 1, & s' = s\\
         0, & \text{else}\end{array}\right. .$
\item $\mathsf{R}$: $\mathsf{R}(r|s,a) = \left\{\begin{array}{ll} \frac{\lVert\mathbf{w}_{T-1}^H \mathbf{h}\rVert_2^2}{\lVert\mathbf{h}\rVert_2^2}, & t = T-1\\
         0, & \text{else}\end{array}\right. .$
\item $\Omega$: The set of possible (noisy) observations $y$: $\Omega \subseteq \mathbb{C}$
\item $\mathsf{O}$: The set of conditional observation probabilities: $y \sim O(y|s,a,n)$
\item $\gamma$: Set to $1$, i.e., an undiscounted \ac{POMDP}
\end{itemize}
Note that the states of this \ac{POMDP} are implicitly given with $\mathbf{h}$, and the actions $a$ with $\mathbf{w}$. Also, in order to use standard \ac{NN} frameworks, the complex valued observations $y$ are converted to real numbers by interpreting them as $2$ dimensional real valued vectors.\\
The agent-environment interaction for this \ac{POMDP} is also shown in Fig. \ref{fig:agent_environment_interaction_ba}.
Note, that the reward function results in a reward signal which can be regarded as being sparse and, thus, renders the task more difficult to learn for a \ac{RL} agent \cite{reward1}. Improving this reward function using reward shaping by introducing a potential-based shaping function \cite{reward2} (e.g. the information of the agent about the \ac{CSI}) for example may further improve performance and is left for future research.\\
Now, the last thing open to define is the agent's action space $\mathsf{A}$ and the mapping of its actions $\mathbf{a}_t$ to the respective combining vectors $\mathbf{w}_t$, i.e., determining a possibly highly nonlinear function 
\begin{equation}
    \mathbf{w}_t = \mathbf{f}(\mathbf{a}_t) \, \in \mathbb{C}^{N_{\text{RX}}}
\end{equation}
and $\mathbf{f}: \mathsf{A} \rightarrow \mathsf{W}_{\text{feasible}}$ with $\mathsf{A} \subseteq \mathbb{R}^{N_{\mathsf{A}}}$, action space dimension $N_{\mathsf{A}}$ and $\mathsf{W}_{\text{feasible}} \subseteq \mathbb{C}^{N_{\text{RX}}}$ (also depicted in Fig. \ref{fig:system_model_ba}).
As the complex-valued combining vectors $\mathbf{w}_t$ often have to obey certain constraints (unit norm and/or unit modulus constraint in fully analogue beamforming, for example), finding the optimal map $\mathbf{f}$ from the real valued action space of the agent to the complex space of feasible combining vectors $\mathsf{W}_{\text{feasible}}$ is not an easy task. Since the action space directly influences the difficulty and complexity of the learning task for the \ac{DRL} agent, ideally one wants to keep the action space and its structure as simple as possible. On the other hand, there is a certain subset of (largely unknown) required combining vectors $ \mathsf{W}_{\text{required}} \subseteq \mathsf{W}_{\text{feasible}}$ which are necessary for any \ac{BA} algorithm to perform well, and $\mathbf{f}$ needs to cover/parameterize (like highly directional beams to achieve high beamforming gain). This greatly increases the difficulty of designing $\mathbf{f}$ as one essentially needs to strike a balance of a simple structure of $\mathsf{A}$ while ensuring good coverage of $\mathsf{W}_{\text{required}}$. In the following, two different implementations for $\mathbf{f}$ under the constraint that $\left|\left|\mathbf{w}\right|\right|_2 =1$ are proposed and described in more detail. Note that for many cases, both implementations can be easily extended to consider more severe and realistic constraints on $\mathbf{w}$.
\subsection{Direct Map}
\label{subsec:direct}
This map describes the direct mapping of the actions of the agent to the combining vectors. Here, the actions $\mathbf{a}_t$ of the agent  at each timestep $t$, are assumed to be real-valued $2N_{\text{RX}}$-dimensional vectors, i.e., $\mathbf{a}_t \in \mathbb{R}^{2N_{\text{RX}}}$ and can be seen as the unscaled, real-valued representation of the corresponding normalized combining vector $\mathbf{w}_t$. Therefore, these actions are then mapped to the complex domain via
\begin{equation}
    \mathbf{w}_{\text{unscaled},t} = (\mathbf{a}_{[0:N_{\text{RX}}],t} +j\mathbf{a}_{[N_{\text{RX}}:2N_{\text{RX}}],t}) \in \mathbb{C}^{N_{\text{RX}}}
\end{equation}
followed by normalization, i.e.,
\begin{equation}
    \mathbf{w}_t = \frac{\mathbf{w}_{\text{unscaled},t}}{\lVert\mathbf{w}_{\text{unscaled},t}\rVert_2}
\end{equation}
with the notation $\mathbf{x}_{[0:R]}$ representing the $R$-dimensional vector consisting of the the first $R$ elements of $\mathbf{x}$.\\
The advantage of this mapping lies in its ability to reach all possible combining vectors, i.e., it covers all of $\mathsf{W}_{\text{feasible}}$. This however comes at the cost of a highly complex structure of $\mathsf{A}$ together with a linear scaling of $N_\mathsf{A}$ by the number of antenna elements $N_{\text{RX}}$. As shown in the results section, these drawbacks seem to pose a challenge too great to overcome for the \ac{DRL} agent, causing it to fail to learn when the problem sizes reach practically relevant dimensions.
\subsection{Beamforming Map}
\begin{figure}[t]
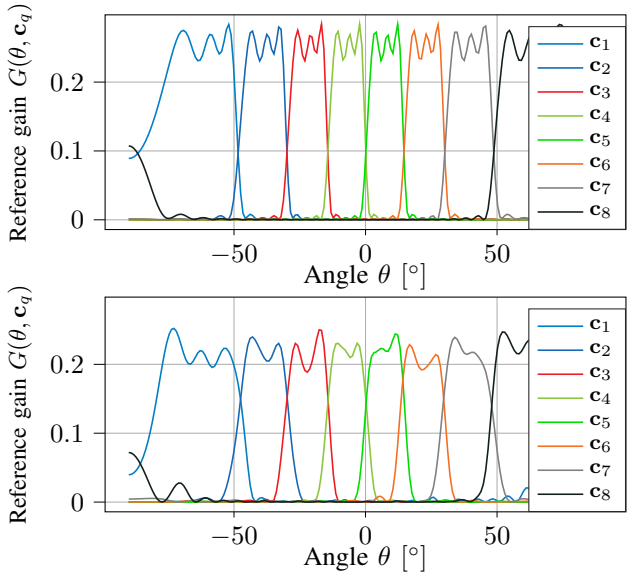

	\centering
	\vspace{-0.5cm}
\begin{subfigure}[b]{0.55\textwidth}
\input{tikz/beamforming_module_1.tikz}
   \label{fig:1} 
   \vspace{-0.3cm}
\end{subfigure}
\begin{subfigure}[b]{0.55\textwidth}
  \input{tikz/beamforming_module_2.tikz}
   \label{fig:2}
\end{subfigure}
\caption{\small Top: Equal beam width codebook with $8$ entries generated by the algorithm described in \cite{codebook}, as reference. Bottom: Beampatterns with the same properties generated using the proposed beamforming map.}
	\label{fig:beamforming_module}
	\vspace{-0.3cm}
\end{figure}

The second map (in the following referred to as "beamforming map" or "beamforming module") tries to address the problems of the direct map in the following way: The foundation is the observation that \ac{BA} algorithms employing hierarchical codebooks show good performance \cite{Codebook_2014} \cite{ActiveLearning}, i.e., covering much of $\mathsf{W}_{\text{required}}$, while the beams of these codebooks are well defined by their angular direction $\alpha_{\text{beam}} \in [-\frac{\pi}{2}, \frac{\pi}{2}]$ and angular beam width $2 \cdot \beta_{\text{beam}}$ with $\beta_{\text{beam}} \in (0,\frac{\pi}{2}]$, i.e., by only two real valued parameters. Thus, the idea is to generate these beampatterns (specified by $\alpha_{\text{beam}}$ and $\beta_{\text{beam}}$) 
``on the fly", i.e., at each timestep $t$ of the \ac{BA} algorithm, the agent inputs a specific $\alpha_{\text{beam},t}$ and $\beta_{\text{beam},t}$ into the beamforming module which then returns the beampattern ($\mathbf{w}_t$) specified by these inputs, thereby drastically reducing $N_\mathsf{A}$ from $2N_{\text{RX}}$ to $2$. As the generation algorithms for the respective codebooks however require extensive and lengthy numerical optimization for each beampattern, the continuation of this idea is to skip this online optimization process by using a pretrained \ac{DNN} instead, resulting in a parameterized mapping with weights $ \boldsymbol{\Theta} $, i.e., $\mathbf{f}$ is now a parameterized function $\mathbf{f}_{\Theta}$ and %
\begin{equation}
    \mathbf{w} = \mathbf{f}_{\boldsymbol{\Theta}} (\alpha_{\text{beam}},\beta_{\text{beam}})
\end{equation}
This \ac{DNN} representing $\mathbf{f}_{\boldsymbol{\Theta}}$ is a simple feedforward \ac{NN} with $2$ input units and $2N_{\text{RX}}$ output units. In the following, it is implicitly assumed that the $2N_{\text{RX}}$ dimensional real valued output of the \ac{DNN} is mapped to normalized complex combining vectors as in \ref{subsec:direct} and that this step is part of the beamforming module.
From a high-level perspective, training of this \ac{DNN} can be done by inputting random input pairs ($\alpha_{\text{beam}},\beta_{\text{beam}}$) into the \ac{DNN} and then adjusting its weights such that its produced beampatterns match the specification of its input as close as possible, e.g. by maximizing the received power inside the angular range $[\alpha_{\text{beam}}-\beta_{\text{beam}},\alpha_{\text{beam}}+\beta_{\text{beam}}]$ and minimizing it outside. Now, a more detailed description follows.\\
A random input pair $(\alpha_{\text{beam}}, \beta_{\text{beam}})$, $\alpha_{\text{beam}} \sim \mathcal{U}[- \frac{\pi}{2}, \frac{\pi}{2}]$ and $\beta_{\text{beam}} \sim (0, \beta_{\text{max}}]$ with $ \beta_{\text{max}} = \text{min}(\frac{\pi}{2},|\frac{\pi}{2}-|\alpha_{\text{beam}}||)$ is drawn uniformly. Then, the following is done: First, the intervals $U_{\text{inside}} = [-\pi, \pi] \cup [2(\alpha_{\text{beam}}-\beta_{\text{beam}}), 2(\alpha_{\text{beam}}+\beta_{\text{beam}})$] and  $U_{\text{outside}} = [-\pi, \pi] \backslash U_{\text{inside}}$ are determined. After this, $K$ random angles are uniformly sampled from $U_{\text{inside}}$ and $U_{\text{outside}}$ respectively (resulting in the sets $\mathcal{S}_{\text{inside}}$ and $\mathcal{S}_{\text{outside}}$), i.e., sampling uniformly in the $\Psi$ - space \cite{codebook}. Then, $(\alpha_{\text{beam}}, \beta_{\text{beam}})$ is fed into the \ac{DNN} and the combining vector $\mathbf{w}$ is obtained from its output. Following this, the weights of the \ac{DNN} are optimized via stochastic gradient descent (averaged over $B$ input samples) to minimize the following objective:
\begin{equation}
   J(\boldsymbol{\Theta}) = -\mathbb{E}_{\text{inside}}[g] + \mathbb{E}_{\text{outside}}[g] + \mathrm{Var}_{\text{outside}}(g) + \epsilon \cdot \mathrm{Var}_{\text{inside}}(g)
   \label{eq:beamforming_module_training}
\end{equation}
with $g = \lVert \mathbf{w}^H\mathbf{a}_{\Psi}(\theta)\rVert_2$ ($\mathbf{a}_{\Psi}(\phi)$ denotes a modified (\ref{eq:antenna_array_response}) where the $\pi \sin(\phi)$-terms are replaced with $\phi$) and with the subscripts \emph{inside} and \emph{outside} indicating that the expectation is taken over all angles $\theta$ from $\mathcal{S}_{\text{inside}}$ and $\mathcal{S}_{\text{outside}}$ respectively. The variance terms in (\ref{eq:beamforming_module_training}) are used as regularizers to generate more smooth beampatters with $\epsilon \geq 0$ being a hyperparameter to control the ripple inside the beampattern's angular coverage range.
Fig \ref{fig:beamforming_module} (bottom) shows the resulting beampatterns when generating a set of $8$ beams similar to a simple equal beam-width codebook with $8$ entries, using a trained \ac{DNN} beamforming module, assuming $N_{\text{RX}} = 32$, $\text{batch size} = 1000$, $K=1000$, $\epsilon = 1$, $5000$ parameter updates and using a $3$ Layer \ac{DNN} with $128$ units in the first two layers ($25 \text{K}$ parameters in total). Plotted in Fig. \ref{fig:beamforming_module} is the reference gain, defined as \cite{codebook}:
\begin{equation}
    G(\theta,\mathbf{c}_q) = \lVert \mathbf{c}_q^H\mathbf{a}(\theta)\rVert_2^2
\end{equation}
over all possible \ac{AoA} $ \theta \in [-\frac{\pi}{2},\frac{\pi}{2}]$ and where each codeword $\mathbf{c}_q$ represents a specific combining vector $\mathbf{w}_q$ (note that here the index $q$ refers to the index in the codeword and not a timestep).
For example, $\mathbf{c}_4$ in Fig. \ref{fig:beamforming_module} (bottom) was generated by the beamforming module by inputting $\alpha_{\text{beam}}=-11.25$ and $\beta_{\text{beam}} = 22.5$.
For comparison, the top of the figure displays beampatterns from a codebook with $8$ entries generated with a classical numerical optimization approach described in \cite{codebook} with algorithm parameters $N =3 $, $M=12$,  $L=300$. While the beampatterns of the beamforming map are not as sharp as the ones generated using the algorithm of \cite{codebook}, they can be generated "on-the-fly" for arbitrary input combinations since much of the complexity of beampattern generation is offloaded to the training phase of the \ac{DNN}. Also, the beamforming map is inherently differentiable due to its \ac{NN}-based structure, meaning one can exchange the feedforward-\ac{NN} used in the approach of \cite{RNNE2E} with this pretrained beamforming map and freeze its parameters in order to gain insights about the performance limitations induced by the beamforming module. Further, one can extend this idea in a straightforward fashion to other schemes, i.e., generating beams with multiple directions, and to other and different hardware restrictions.
\section{Numerical Results}
\begin{table}
	\caption{\small Hyperparameters for architecture and training of the \ac{DRL} agent}
	\label{tab:hyperparameter}
	\centering
	\begin{tabular}{l|l}
		Parameter& Value \\
		\hline
		Units per Layer & 128\\
		GRU Layers & 2\\
		FF Layers & 2 \\
		Entropy Coefficient & 0.001 \\
		Clip Coefficient & 0.2 \\
		$\gamma$ & 1\\
	\end{tabular}
	\begin{tabular}{l|l}
		Parameter& Value \\
		\hline
		Max. grad norm & $0.5$ \\
		Training SNR  & $20$ dB  \\
		Batch size & $2000$  \\
		Optimizer & ADAM \\
		Learning rate & $3 \cdot 10^{-4}$ \\
		Value Coefficient & 0.5\\
		Number of workers & 2000
	\end{tabular}
	\vspace{-0.5cm}
\end{table}

\label{sec:result}
In this section we present numerical results for the \ac{DRL} agent using the two different beamforming maps previously introduced (in the following denoted as $\text{DRL}_{\text{DM}}$ and $\text{DRL}_{\text{BF}}$). In all experiments, the \ac{DRL} agent was trained over the full-length rollouts of the \ac{BA} algorithm, i.e., $T$ timesteps, using \ac{BPTT}. Table \ref{tab:hyperparameter} lists the most important hyperparameters used for our experiments. Also, the number of parameters of all chosen \ac{NN}-based approaches were selected to be roughly equal for easier comparison. We compare our results to the following five baselines:
\begin{enumerate}
    \item Maximum ratio combining (MRC) \acused{MRC} with perfect \ac{CSI} ($\text{MRC}_{\text{CSI}})$: Under the assumption of full \ac{CSI} knowledge at the receiver, this can be seen as the optimal scheme and thus represents an upper bound on the achievable beamforming gain.
    \item MRC with estimated \ac{CSI} ($\text{MRC}_{\text{OMP}}$)\footnote{The implementation of this approach was taken from the github repository of \cite{RNNE2E}.}: Here, the \ac{CSI} for the \ac{MRC} scheme is estimated using compressed sensing (\ac{OMP}).
    \item \ac{RNN} based unsupervised \cite{RNNE2E} ($\text{RNN}_{\text{UNSUP}})$: This approach trains a \ac{RNN} in a unsupervised end-to-end fashion to maximize the beamforming gain at the last timestep $T-1$.
    \item \ac{DNN} non-adaptive unsupervised \cite{9367586} ($\text{DNN}_{\text{UNSUP}}$): In this scheme, a \ac{DNN} is trained to maximize the beamforming gain at the last timestep $T-1$ after observing $T-1$ pilots obtained by sensing the channel using a sequence of fixed, random sensing vectors.
    \item Exhaustive search based on a codebook generated as in \cite{codebook}.
\end{enumerate}
Note, that both $\text{RNN}_{\text{UNSUP}}$ and $\text{DNN}_{\text{UNSUP}}$ are trained in an unsupervised end-to-end fashion by maximizing the beamforming gain, requiring an explicit channel model for optimization during training, which our approach does not need.
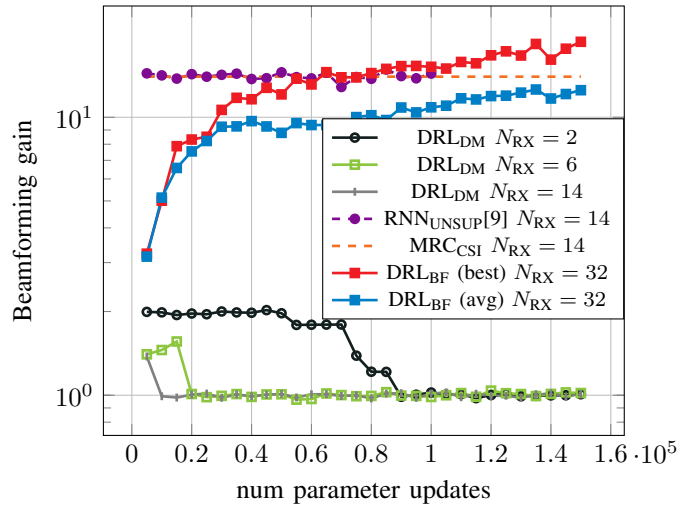
\begin{figure}[t]
	\centering
	\input{tikz/direct_map_training.tikz}
	\vspace{-0.4cm}
	\caption{\small Beamforming gain during training of the $\text{DRL}_{\text{DM}}$ agent using the direct map for different values of $N_{\text{RX}}$. Also plotted is the training trajectory of a $\text{DRL}_{\text{BF}}$ agent with $N_{\text{RX}} = 32$ antennas, exhibiting a much more favorable training curve than the $\text{DRL}_{\text{DM}}$ approaches even though it uses more antenna elements. \emph{(avg)} denotes the average taken over $5$ identical agents trained with different random seeds and 
\emph{(best)} denotes the best performing one out of these $5$ agents.}
	\label{fig:direct_map_training}
	\vspace{-0.75cm}
\end{figure}
First, the \ac{DRL} approach using the direct map ($\text{DRL}_{\text{DM}}$) is investigated. For this, a separate $\text{DRL}_{\text{DM}}$ agent is trained for each different value of $N_{\text{RX}}$, starting from $2$ up to $14$ with the running time $T$ of the algorithm kept constant to $T=5$, $L=1$, with a total of $100 \text{K}$ parameter updates. The results can be seen in Fig. \ref{fig:direct_map_training}: Whereas the agent more or less successfully manages to learn a policy for the $2$ antenna case, even moderately increasing the amount of antennas quickly causes the learning to fail. Also note the loss of convergence during training which may be mitigable through early stopping or further hyperparameter tuning. Although some light hyperparameter tuning for the \ac{DRL} agent was attempted without much success, making definite statements about the performance of \ac{DRL} agents in general is not an easy task, as also outlined in \cite{DRLmatters}. However, given how quickly the performance of the agent declined when increasing $N_{\text{RX}}$, the chances that one obtains good and working hyperparameters for the direct map for realistic problem sizes seem very slim.

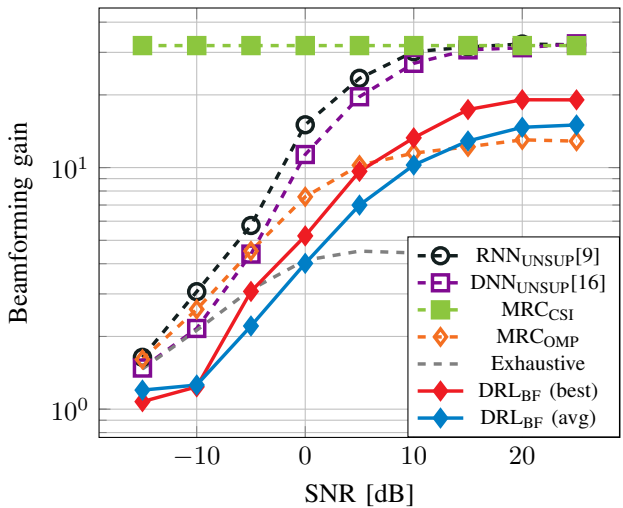
\begin{figure}[t]
	\centering
	\input{tikz/results/performance_comparison.tikz}
	\vspace{-0.1cm}
	\caption{\small Beamforming gain vs. SNR of the $\text{DRL}_{\text{BF}}$ agent and various other methods for $T=5$, $N_{\text{RX}}=32$. Note that as expected, the $\text{DRL}_{\text{BF}}$ agent fails to match the performance of its unsupervised counterparts, but it manages to outperform the $\text{MRC}_{\text{OMP}}$ in the high \ac{SNR} regime.} 
	\label{fig:performance_comparison}
	\vspace{-0.6cm}
\end{figure}

Next, a new experiment is performed in which the performance of the \ac{DRL} agent using the beamforming map ($\text{DRL}_{\text{BF}}$) is more closely examined and compared to other approaches. For this, a trained $3$-layer ($25$K parameters) \ac{DNN} beamforming map is used and all hyperparameters of the \acp{NN} are kept the same as in the previous experiment, $T=5$ and $N_{\text{RX}}=32$. The results are shown in Fig. \ref{fig:performance_comparison}. First one can observe that using this beamforming map, the \ac{DRL} agent is able to learn a \ac{BA} policy for $N_{\text{RX}} = 32$, which is quite remarkable in itself given the previous results using the direct map (the first $150$K parameter updates during training are also shown in Fig. \ref{fig:direct_map_training}). Although the $\text{DRL}_{\text{BF}}$ agent fails to reach the performance of its unsupervised end-to-end counterparts and only manages to slightly outperform the $\text{MRC}_{\text{OMP}}$ approach in the high \ac{SNR} regime, with the help of better beamforming modules, more hyperparameter optimization and other advanced \ac{DRL} techniques, further possible performance enhancements to the $\text{DRL}_{\text{BF}}$ agent should be possible. We also want to note that addressing the sample efficiency of the \ac{DRL} agent was not the focus of the present study and is left for future investigations.
\section{Conclusion}
\label{sec:conclusion}
This work demonstrates the possibility of tackling the problem of adaptive initial access beam alignment in mmWave communications using a \ac{DRL} framework and the state-of-the-art \ac{PPO} algorithm. \ac{DRL} based methods have a theoretical advantage over current deep unsupervised end-to-end methods designed for this problem in that they do not require any form of environment model and can learn just from interactions with the environment. It is shown that while the direct application of the off-the-shelf \ac{PPO} algorithm fails on relevant problem sizes, using a specially designed beamforming module enables it to scale and deliver competitive performance, even under realistic problem sizes. Due to its \ac{DNN}-based implementation, the proposed beamforming module can be easily trained and adapted to different antenna configurations and hardware constraints, providing a large amount of flexibility. There are multiple possible future directions: one can for example study more and possibly better
beamforming modules as they have been shown to have a great
impact on performance, or, investigate the performance when
using a frame-stack method in favour of a memory network for the \ac{DRL} agent.

\bibliographystyle{IEEEtran}
\bibliography{IEEEabrv,references}

\end{document}

%% file: acronyms.tex
\begin{acronym}
 \acro{CSI}{channel state information}
 \acro{UE}{user equipment}
 \acro{UL}{uplink}
 \acro{BS}{basestation}
 \acro{TDD}{time division duplex}
 \acro{FDD}{frequency division duplex}
 \acro{ECC}{error-correcting code}
 \acro{MLD}{maximum likelihood decoding}
 \acro{HDD}{hard decision decoding}
 \acro{IF}{intermediate frequency}
 \acro{RF}{radio frequency}
 \acro{SDD}{soft decision decoding}
 \acro{NND}{neural network decoding}
 \acro{CNN}{convolutional neural network}
 \acro{ML}{maximum likelihood}
 \acro{GPU}{graphical processing unit}
 \acro{BP}{belief propagation}
 \acro{LTE}{Long Term Evolution}
 \acro{BER}{bit error rate}
 \acro{DER}{detection error rate}
 \acro{SNR}{signal-to-noise-ratio}
 \acro{ReLU}{rectified linear unit}
 \acro{BPSK}{binary phase shift keying}
 \acro{QPSK}{quadrature phase shift keying}
 \acro{AWGN}{additive white Gaussian noise}
 \acro{MSE}{mean squared error}
 \acro{LLR}{log-likelihood ratio}
 \acro{MAP}{maximum a posteriori}
 \acro{NVE}{normalized validation error}
 \acro{BCE}{binary cross-entropy}
 \acro{CE}{cross-entropy}
 \acro{BLER}{block error rate}
 \acro{SQR}{signal-to-quantisation-noise-ratio}
 \acro{MIMO}{multiple-input multiple-output}
 \acro{OFDM}{orthogonal frequency division multiplex}
 \acro{RF}{radio frequency}
 \acro{LOS}{line of sight}
 \acro{NLoS}{non-line of sight}
 \acro{NMSE}{normalized mean squared error}
 \acro{CFO}{carrier frequency offset}
 \acro{SFO}{sampling frequency offset}
 \acro{IPS}{indoor positioning system}
 \acro{TRIPS}{time-reversal IPS}
 \acro{RSSI}{received signal strength indicator}
 \acro{MIMO}{multiple-input multiple-output}
 \acro{ENoB}{effective number of bits}
 \acro{AGC}{automated gain control}
 \acro{ADC}{analog to digital converter}
 \acro{ADCs}{analog to digital converters}
 \acro{FB}{front bandpass}
 \acro{FPGA}{field programmable gate array}
 \acro{JSDM}{Joint Spatial Division and Multiplexing}
 \acro{NN}{neural network}
 \acro{IF}{intermediate frequency}
 \acro{LoS}{line-of-sight}
 \acro{NLoS}{non-line-of-sight}
 \acro{DSP}{digital signal processing}
 \acro{AFE}{analog front end}
 \acro{SQNR}{signal-to-quantisation-noise-ratio}
 \acro{SINR}{signal-to-interference-noise-ratio}
 \acro{ENoB}{effective number of bits}
 \acro{AGC}{automated gain control}
 \acro{PCB}{printed circuit board}
 \acro{EVM}{error vector mangnitude}
 \acro{CDF}{cumulative distribution function}
 \acro{MRC}{maximum ratio combining}
 \acro{MRP}{maximum ratio precoding}
 \acro{MRT}{maximum ratio transmission}
 \acro{DeepL}{deep-learning}
 \acro{DL}{deep learning}
 \acro{SISO}{single-input single-output}
 \acro{SGD}{stochastic gradient descent}
 \acro{CP}{cyclic prefix}
 \acro{MISO}{Multiple Input Single Output}
 \acro{LMMSE}{linear minimum mean square error}
 \acro{ZF}{zero forcing}
 \acro{USRP}{universal software radio peripheral}
 \acro{RNN}{recurrent neural network}
 \acro{GRU}{gated recurrent unit}
 \acro{LSTM}{long short-term memory}
 \acro{NTM}{neural turing machine}
 \acro{DNC}{differentiable neural computer}
 \acro{TCN}{temporal convolutional network}
 \acro{FCL}{fully connected layer}
 \acro{MANN}{memory augmented neural network}
 \acro{RNN}{recurrent neural network}
 \acro{DNN}{dense neural network}
 \acro{FIR}{finite impulse response}
 \acro{BPTT}{back-propagation through time}
 \acro{GAN}{generative adversarial network}
 \acro{ELU}{exponential linear unit}
 \acro{tanh}{hyperbolic tangent}
 \acro{BICM}{bit-interleaved coded modulation}
 \acro{OTA}{over-the-air}
 \acro{IM}{intensity modulation}
 \acro{DD}{direct detection}
 \acro{RL}{reinforcement learning}
 \acro{SDR}{software-defined radio}
 \acro{WGAN}{Wasserstein generative adversarial network}
 \acro{BMD}{bit-metric decoding}
 \acro{BMI}{bit-wise mutual information}
 \acro{LDPC}{low-density parity-check}
 \acro{IDD}{iterative demapping and decoding}
 \acro{IEDD}{iterative equalization, demapping and decoding}
 \acro{JSD}{Jensen-Shannon divergence}
 \acro{MMSE}{minimum mean square error}
 \acro{FFT}{fast Fourier transform}
 \acro{IFFT}{inverse fast Fourier transform}
 \acro{QAM}{quadrature amplitude modulation}
 \acro{EMD}{earth mover's distance}
 \acro{TDL}{tapped delay line}
 \acro{KL}{Kullback-Leibler}
 \acro{PRACH}{physical random access channel}
 \acro{URLLC}{ultra-reliable low-latency communication}
 \acro{ANOMA}{asynchronous non-orthogonal multiple access}
 \acro{FEC}{forward error correction}
 \acro{NOMA}{non-orthogonal medium access}
 \acro{MTC}{machine-type communications}
 \acro{mMTC}{massive machine-type communications}
 \acro{MCS}{modulation and coding scheme}
 \acro{PAPR}{peak-to-average power ratio}
 \acro{MAC}{medium access control}
 \acro{STO}{sampling time offset}
 \acro{STE}{straight-through estimator}
 \acro{PHY}{physical}
 \acro{CCE}{categorical cross-entropy}
 \acro{IoT}{Internet of Things}
 \acro{CCDF}{complementary cumulative distribution function}
 \acro{CRC}{cyclic redundancy check}
 \acro{ACLR}{adjacent channel leakage ratio}
\acro{MD}{missed detection}
\acro{FA}{false alarm}
\acro{FAR}{false alarm rate}
\acro{BCJR}{Bahl-Cocke-Jelinek-Raviv}
\acro{BA}{beam alignment}
\acro{DRL}{deep reinforcement learning}
\acro{MDP}{markov decision process}
\acro{POMDP}{partially observable Markov decision process}
\acro{ULA}{uniform linar array}
\acro{PPO}{proximal policy optimization}
\acro{AoA}{angle-of-arrival}
\acro{IA}{initial access}
\acro{OMP}{orthogonal matching pursuit}
\end{acronym}

%% file: corporateColours.tex
\definecolor{mittelblau}{RGB}{0, 126, 198}
\definecolor{violettblau}{cmyk}{0.9, 0.6, 0, 0}
\definecolor{rot}{RGB}{238, 28 35}
\definecolor{apfelgruen}{RGB}{140, 198, 62}
\definecolor{gelb}{RGB}{1, 221, 0}
\definecolor{orange}{RGB}{244, 111, 33}
\definecolor{pink}{RGB}{237, 0, 140}
\definecolor{lila}{RGB}{128, 10, 145}
\definecolor{hellgrau}{RGB}{224, 224, 224}
\definecolor{mittelgrau}{RGB}{128, 128, 128}
\definecolor{dunkelgrau}{RGB}{80,80,80}
\definecolor{anthrazit}{RGB}{19, 31, 31}

%% file: plotcyclelists.tex
\pgfplotscreateplotcyclelist{corporate colours markers}{%
rot, every mark/.append style={fill=.!80!rot},mark=*\\%
mittelblau, every mark/.append style={fill=.!80!mittelblau},mark=square*\\%
apfelgruen, every mark/.append style={fill=.!80!apfelgruen},mark=triangle*\\%
orange, mark=star\\%
pink, every mark/.append style={fill=.!80!pink},mark=diamond*\\%
violettblau, every mark/.append style={fill=.!80!violettblau},mark=otimes*\\%
lila, mark=|\\%
gelb, every mark/.append style={fill=.!80!gelb},mark=pentagon*\\%
hellgrau, mark=text,text mark=p\\%
anthrazit, mark=text,text mark=a\\%
}

\pgfplotscreateplotcyclelist{corporate colours markers double}{%
rot, every mark/.append style={fill=.!80!rot},mark=*\\%
rot, dashed, every mark/.append style={fill=.!80!rot,solid},mark=*\\%
mittelblau, every mark/.append style={fill=.!80!mittelblau},mark=square*\\%
mittelblau, dashed, every mark/.append style={fill=.!80!mittelblau,solid},mark=square*\\%
apfelgruen, every mark/.append style={fill=.!80!apfelgruen},mark=triangle*\\%
apfelgruen, dashed, every mark/.append style={fill=.!80!apfelgruen,solid},mark=triangle*\\%
orange, mark=star\\%
orange, dashed, mark=star\\%
pink, every mark/.append style={fill=.!80!pink},mark=diamond*\\%
pink, dashed, every mark/.append style={fill=.!80!pink,solid},mark=diamond*\\%
violettblau, every mark/.append style={fill=.!80!violettblau},mark=otimes*\\%
violettblau, dashed, every mark/.append style={fill=.!80!violettblau,solid},mark=otimes*\\%
lila, mark=|\\%
lila, dashed, mark=|\\%
gelb, every mark/.append style={fill=.!80!gelb},mark=pentagon*\\%
gelb, dashed, every mark/.append style={fill=.!80!gelb,solid},mark=pentagon*\\%
}

\pgfplotsset{
  colormap/magma/.style={%
    /pgfplots/colormap={magma}{%
      rgb=(0.001462, 0.000466, 0.013866)
      rgb=(0.035520, 0.028397, 0.125209)
      rgb=(0.102815, 0.063010, 0.257854)
      rgb=(0.191460, 0.064818, 0.396152)
      rgb=(0.291366, 0.064553, 0.475462)
      rgb=(0.384299, 0.097855, 0.501002)
      rgb=(0.475780, 0.134577, 0.507921)
      rgb=(0.569172, 0.167454, 0.504105)
      rgb=(0.664915, 0.198075, 0.488836)
      rgb=(0.761077, 0.231214, 0.460162)
      rgb=(0.852126, 0.276106, 0.418573)
      rgb=(0.925937, 0.346844, 0.374959)
      rgb=(0.969680, 0.446936, 0.360311)
      rgb=(0.989363, 0.557873, 0.391671)
      rgb=(0.996580, 0.668256, 0.456192)
      rgb=(0.996727, 0.776795, 0.541039)
      rgb=(0.992440, 0.884330, 0.640099)
      rgb=(0.987053, 0.991438, 0.749504)
    },
  },
  colormap/inferno/.style={%
    /pgfplots/colormap={inferno}{%
      rgb=(0.001462, 0.000466, 0.013866)
      rgb=(0.037668, 0.025921, 0.132232)
      rgb=(0.116656, 0.047574, 0.272321)
      rgb=(0.217949, 0.036615, 0.383522)
      rgb=(0.316282, 0.053490, 0.425116)
      rgb=(0.410113, 0.087896, 0.433098)
      rgb=(0.503493, 0.121575, 0.423356)
      rgb=(0.596940, 0.154848, 0.398125)
      rgb=(0.688653, 0.192239, 0.357603)
      rgb=(0.775059, 0.239667, 0.303526)
      rgb=(0.851384, 0.302260, 0.239636)
      rgb=(0.912966, 0.381636, 0.169755)
      rgb=(0.956852, 0.475356, 0.094695)
      rgb=(0.981895, 0.579392, 0.026250)
      rgb=(0.987464, 0.690366, 0.079990)
      rgb=(0.973088, 0.805409, 0.216877)
      rgb=(0.947594, 0.917399, 0.410665)
      rgb=(0.988362, 0.998364, 0.644924)
    },
  },
  colormap/plasma/.style={%
    /pgfplots/colormap={plasma}{%
      rgb=(0.050383, 0.029803, 0.527975)
      rgb=(0.186213, 0.018803, 0.587228)
      rgb=(0.287076, 0.010855, 0.627295)
      rgb=(0.381047, 0.001814, 0.653068)
      rgb=(0.471457, 0.005678, 0.659897)
      rgb=(0.557243, 0.047331, 0.643443)
      rgb=(0.636008, 0.112092, 0.605205)
      rgb=(0.706178, 0.178437, 0.553657)
      rgb=(0.768090, 0.244817, 0.498465)
      rgb=(0.823132, 0.311261, 0.444806)
      rgb=(0.872303, 0.378774, 0.393355)
      rgb=(0.915471, 0.448807, 0.342890)
      rgb=(0.951344, 0.522850, 0.292275)
      rgb=(0.977856, 0.602051, 0.241387)
      rgb=(0.992541, 0.687030, 0.192170)
      rgb=(0.992505, 0.777967, 0.152855)
      rgb=(0.974443, 0.874622, 0.144061)
      rgb=(0.940015, 0.975158, 0.131326)
    },
  },
}

%% file: system.tikzstyles
\tikzstyle{test 1}=[fill=white, draw=black, shape=circle]
\tikzstyle{plus}=[fill=white, draw=black, shape=circle, scale=0.5]

\tikzstyle{arrow 1}=[->, -stealth, draw={rgb,255: red,62; green,48; blue,255}, fill={rgb,255: red,62; green,48; blue,255}, ultra thick]
\tikzstyle{box 1}=[-, draw={blue}, fill opacity=0.2, ultra thick]
\tikzstyle{box 2}=[-, draw={orange}, fill opacity=0.2, ultra thick]
\tikzstyle{box 3}=[-, fill={red}, draw=red, fill opacity=0.2, ultra thick]
\tikzstyle{arrow 2}=[->, -stealth, semithick]
\tikzstyle{box3}=[-, fill=white]
\tikzstyle{arrow 3}=[->, -stealth, semithick, double]
\tikzstyle{arrow 4}=[->, -stealth, thick]
\tikzstyle{arrow 5}=[-, thick]
\tikzstyle{arrow 6}=[->, -stealth, thick, double]
\tikzstyle{arrow 7}=[-, thick, double]
\tikzstyle{arrow 8}=[-, semithick, double]

%% file: tikz/system_model.tikz
\begin{tikzpicture}
 \tikzstyle{every node}=[font=\huge]
\begin{pgfonlayer}{nodelayer}
		\node [style=none] (0) at (0, 0.75) {};
		\node [style=none] (1) at (1.75, 0.75) {};
		\node [style=none] (2) at (1.75, -0.25) {};
		\node [style=none] (3) at (0, -0.25) {};
		\node [style=none] (4) at (12, 3.25) {};
		\node [style=none] (5) at (20.5, 3.25) {};
		\node [style=none] (6) at (12, -6.5) {};
		\node [style=none] (7) at (20.5, -6.5) {};
		\node [style=none] (12) at (1.75, 0.25) {};
		\node [style=none] (15) at (15.75, 0) {};
		\node [style=none] (17) at (18.25, -0.75) {};
		\node [style=none] (37) at (10, 2) {};
		\node [style=none] (38) at (10, 2) {};
		\node [style=none] (39) at (10, 3) {};
		\node [style=none] (40) at (9.5, 3.75) {};
		\node [style=none] (41) at (10.5, 3.75) {};
		\node [style=none] (48) at (2.75, 0.25) {};
		\node [style=none] (49) at (2.75, 0.25) {};
		\node [style=none] (50) at (2.75, 1.25) {};
		\node [style=none] (51) at (2.25, 0.75) {};
		\node [style=none] (52) at (3.25, 0.75) {};
		\node [style=none] (54) at (3.75, 0.25) {};
		\node [style=none] (55) at (9.75, 0.25) {};
		\node [style=none] (64) at (14, -4) {};
		\node [style=none] (74) at (7, 1.25) {$\mathbf{h}=\sum_{l=1}^{L} \alpha_l \mathbf{a}(\phi_l)$ };
		\node [style=none] (78) at (0.75, 0.25) {UE};
		\node [style=none] (80) at (18.25, 0) {RFC};
		\node [style=none] (82) at (14.25, -4.8) {$\mathbf{w}_t = \mathbf{f}(\mathbf{a}_t)$};
		\node [style=none] (84) at (16, -8.5) {DRL};
		\node [style=none] (86) at (16, -9.25) {Agent};
		\node [style=none] (87) at (10.75, -3.25) {$N_{\text{RX}}$};
		\node [style=none] (88) at (19, -7.25) {$y_t \in \mathbb{C}$};
		\node [style=none] (89) at (12, -7.25) {$\mathbf{a}_t \in \mathbb{R}^{N_{\mathsf{A}}}$};
		\node [style=none] (90) at (16, -3) {$\mathbf{w}_t \in \mathbb{C}^{N_{\text{RX}}}$};
		\node [style=none] (92) at (14, 0.75) {Combiner};
		\node [style=none] (93) at (14, -0.25) {$\mathbf{w}_t \in \mathbb{C}^{N_{\text{RX}}}$};
		\node [style=none] (94) at (12.25, 2.25) {};
		\node [style=none] (95) at (15.75, 2.25) {};
		\node [style=none] (96) at (12.25, -2.25) {};
		\node [style=none] (97) at (15.75, -2.25) {};
		\node [style=none] (98) at (16.5, 0.75) {};
		\node [style=none] (99) at (16.5, -0.75) {};
		\node [style=none] (100) at (20, -0.75) {};
		\node [style=none] (101) at (20, 0.75) {};
		\node [style=none] (102) at (16, -4) {};
		\node [style=none] (103) at (12.25, -4) {};
		\node [style=none] (104) at (12.25, -5.75) {};
		\node [style=none] (105) at (16, -5.75) {};
		\node [style=none] (106) at (19.75, 2.75) {BS};
		\node [style=none] (107) at (7, -0.75) {$\in \mathbb{C}^{N_{\text{RX}}}$};
		\node [style=none] (109) at (14, -2.25) {};
		\node [style=none] (113) at (11, 1) {.};
		\node [style=none] (114) at (11, 0.5) {};
		\node [style=none] (115) at (11, 0.5) {.};
		\node [style=none] (116) at (11, 0) {.};
		\node [style=none] (118) at (16.5, 0) {};
		\node [style=plus] (124) at (11, 2) {+};
		\node [style=none] (125) at (10, -2) {};
		\node [style=none] (126) at (10, -2) {};
		\node [style=none] (127) at (10, -1) {};
		\node [style=none] (128) at (9.5, -0.25) {};
		\node [style=none] (129) at (10.5, -0.25) {};
		\node [style=plus] (130) at (11, -2) {+};
		\node [style=none] (133) at (12.25, 2) {};
		\node [style=none] (134) at (12.25, -2) {};
		\node [style=none] (135) at (11, 3) {};
		\node [style=none] (136) at (11, -1) {};
		\node [style=none] (137) at (14, -3) {};
		\node [style=none] (138) at (11.25, -0.75) {$n_{N_{\text{RX}}}$};
		\node [style=none] (139) at (11.25, 3.25) {$n_1$};
		\node [style=none] (140) at (14, -5.75) {};
		\node [style=none] (141) at (18.25, -5.5) {};
		\node [style=none] (142) at (17.5, -5.5) {};
		\node [style=none] (143) at (17.5, -8) {};
		\node [style=none] (144) at (13.5, -8) {};
		\node [style=none] (145) at (18.5, -8) {};
		\node [style=none] (146) at (13.5, -10) {};
		\node [style=none] (147) at (18.5, -10) {};
		\node [style=none] (148) at (14, -6.75) {};
		\node [style=none] (149) at (14.5, -6.75) {};
		\node [style=none] (150) at (14.5, -8) {};
	\end{pgfonlayer}
	\begin{pgfonlayer}{edgelayer}
		\draw (40.center) to (39.center);
		\draw (39.center) to (41.center);
		\draw (39.center) to (38.center);
		\draw (51.center) to (50.center);
		\draw (50.center) to (52.center);
		\draw (50.center) to (49.center);
		\draw (12.center) to (49.center);
		\draw [style=arrow 1] (54.center) to (55.center);
		\draw [style=box 1] (1.center)
			 to (2.center)
			 to (3.center)
			 to (0.center)
			 to cycle;
		\draw [style=box 2] (4.center)
			 to (5.center)
			 to (7.center)
			 to (6.center)
			 to cycle;
		\draw [style=box3] (97.center)
			 to (96.center)
			 to (94.center)
			 to (95.center);
		\draw [style=box3] (95.center) to (15.center);
		\draw [style=box3] (15.center) to (97.center);
		\draw [style=box3] (99.center)
			 to (98.center)
			 to (101.center)
			 to (100.center)
			 to cycle;
		\draw [style=box3] (103.center)
			 to (102.center)
			 to (105.center)
			 to (104.center)
			 to cycle;
		\draw (38.center) to (124);
		\draw (128.center) to (127.center);
		\draw (127.center) to (129.center);
		\draw (127.center) to (126.center);
		\draw (126.center) to (130);
		\draw [style=arrow 3] (124) to (133.center);
		\draw [style=arrow 3] (130) to (134.center);
		\draw [style=arrow 3] (135.center) to (124);
		\draw [style=arrow 3] (136.center) to (130);
		\draw [style=arrow 7] (64.center) to (137.center);
		\draw [style=arrow 6] (137.center) to (109.center);
		\draw [style=arrow 6] (15.center) to (118.center);
		\draw [style=arrow 8] (17.center) to (141.center);
		\draw [style=arrow 8] (141.center) to (142.center);
		\draw [style=arrow 3] (142.center) to (143.center);
		\draw [style=box 3] (150.center)
			 to (143.center)
			 to (145.center)
			 to (147.center)
			 to (146.center)
			 to (144.center)
			 to cycle;
		\draw [style=arrow 4] (148.center) to (140.center);
		\draw [style=arrow 5] (148.center) to (149.center);
		\draw [style=arrow 5] (149.center) to (150.center);
	\end{pgfonlayer}
\end{tikzpicture}

%% file: tikz/agent_env.tikz
\begin{tikzpicture}[every text node part/.style={align=center}]

\tikzstyle{block} = [rectangle, draw, 
    text width=8em, text centered, rounded corners, minimum height=4em]
    
\tikzstyle{line} = [draw, -latex]
\usetikzlibrary{arrows.meta,shadows,positioning}

\tikzset{
  frame/.style={
    rectangle, draw,
    text width=6em, text centered,
    minimum height=4em,drop shadow,fill=white,
    rounded corners,
  },
  line/.style={
    draw, -{Latex},rounded corners=3mm,
  }
}
\tikzstyle{object}=[rectangle,rounded corners,draw=blue,fill=blue!10,thick,
inner sep=5pt,minimum size=10mm]
  \node (neural_network) [object] { Neural Network: \\ $\mathbf{a}_t = \mathbf{f}(\boldsymbol{\theta},\mathbf{o}_t, \mathbf{h}_{t-1})$ };
  \node (agent) [draw=red, ultra thick, fit={(neural_network)}] {};
  \node[above right, inner sep=0pt, text = red ] at (agent.north west) {DRL Agent};

 \node (symbol) [object,  right = 4cm of agent] { Transmit Symbol:\\ sample $\mathbf{n}_t$\\ $y_t = \mathbf{w}_t^H \mathbf{h} + \mathbf{w}_t^H \mathbf{n}_t$};
\node (process_action) [object, above=of symbol] {Process Action: \\ $\mathbf{w}_t = \mathbf{f}(\mathbf{a}_t)$ };
  \node (prep) [object,below=of symbol] {Calculate Reward and Observation:\\ $r_t = \frac{\lVert \mathbf{w}_t^H \mathbf{h}\rVert_2^2}{\lVert \mathbf{h}\rVert_2^2} $ if $ t = T-1$, $0$ else \\ $\mathbf{o}_{t+1} = y_t$};
 \begin{scope}[blend mode=overlay,overlay]
\node (environment) [draw=orange, ultra thick,fit={(process_action)(symbol)(prep)}] {};
\end{scope}
\node[above left, inner sep=0pt, text = orange ] at (environment.north east) {Environment};
\draw[->, thick] (agent.north) |- ([shift={(0mm,30mm)}]agent.north) node [ right = 2.5 cm,text width=1cm,text centered,midway]{action $\mathbf{a}_t$} -- ([shift={(0mm,7mm)}]environment.north)-| (environment.north);
\draw[->, thick]   ([shift={(0mm,-35mm)}]agent.south) node [ below = 4.25cm ,text width=1cm,text centered,midway]{observation $\mathbf{o}_t$} -- (agent.south);
\draw[->, thick] (environment.260) |- ([shift={(0mm,-5mm)}]environment.260) node [ left = 2 cm,text width=1cm,text centered,midway]{observation $\mathbf{o}_{t+1}$} -- ([shift={(-45mm,-5mm)}]environment.260);
\draw[->, thick] (environment.280) |- ([shift={(0mm,-12mm)}]environment.280) node [ left = 2 cm,text width=1cm,text centered,midway]{reward $r_t$} -- ([shift={(-57mm,-12mm)}]environment.280);
\draw[thick, dashed] ([shift={(-58mm,-14mm)}]environment.280) -- ([shift={(-58mm,-3mm)}]environment.280);
\draw[->, thick] (process_action.south) -- (symbol.north) node [ right,text width=1cm,text centered,midway]{$\mathbf{w}_t$};
\draw[->, thick] (symbol.south) -- (prep.north) node [ right,text width=1cm,text centered,midway]{$\mathbf{w}_t$, $y_t$} ;
\begin{scope}[xscale=-1]
\draw[thick,->, rotate = -20] (neural_network.360) arc (0:-330:5mm);
\end{scope}
\node[right = 1 cm of neural_network]  {$\mathbf{h}_{t-1}$};
\end{tikzpicture}

%% file: tikz/direct_map_training.tikz
\begin{tikzpicture}
		
		\pgfplotsset{compat=1.5}

	\begin{axis}[
	xlabel={num parameter updates},
	ylabel={Beamforming gain},
	xmajorgrids,
	ymode=log,
    every x tick scale label/.style={
    at={(1.01,-0.065)},xshift=1pt,anchor=south west,inner sep=0pt
},
	ymajorgrids,
	legend style={at={(0.42,0.27),font=\tiny}, anchor=south west}]

			\addplot+ [anthrazit, mark=o, mark options={solid},line width=1pt,mark size= 1.5pt] 
			table{
5000 1.995017910003662
10000 1.9867379665374756
15000 1.9435473442077638
20000 1.968033766746521
25000 1.9562895536422729
30000 2.001128816604614
35000 1.9858587741851808
40000 1.9804097890853882
45000 2.023743414878845
50000 1.9721091985702515
55000 1.7904369354248046
60000 1.793121099472046
65000 1.7949193954467773
70000 1.7958845615386962
75000 1.3877258062362672
80000 1.2149402856826783
85000 1.2133693099021912
90000 0.9870276331901551
95000 1.0041778683662415
100000 1.0224146008491517
105000 1.0058080196380614
110000 1.0087260723114013
115000 0.975833785533905
120000 1.001093053817749
125000 1.014052641391754
130000 0.9911065101623535
135000 0.9948864936828613
140000 0.9984571814537049
145000 0.9994000673294068
150000 1.0082000136375426

			};
    	
            \addlegendentry{\footnotesize $\text{DRL}_{\text{DM}}$ $N_{\text{RX}}=2$};
		
			\addplot+ [apfelgruen, mark=square, mark options={solid},line width=1pt,mark size= 1.5pt]
		table{%
5000 1.4033195972442627
10000 1.4541438102722168
15000 1.560726284980774
20000 1.010162079334259
25000 0.984062397480011
30000 0.9948517084121704
35000 1.0095659852027894
40000 0.9861049652099609
45000 1.0066905736923217
50000 1.0077138900756837
55000 0.9640753388404846
60000 0.9691332936286926
65000 1.0145679116249084
70000 1.0012349724769591
75000 0.991854977607727
80000 0.9941167712211609
85000 1.0254098415374755
90000 0.9973010063171387
95000 0.99405277967453
100000 0.9856050252914429
105000 0.9995392799377442
110000 1.0179859399795532
115000 0.9872121572494507
120000 1.039222741127014
125000 1.0169013142585754
130000 1.0098895192146302
135000 0.9930371284484864
140000 1.0107823133468627
145000 1.0236379861831666
150000 1.0183300375938416

			};

            \addlegendentry{\footnotesize $\text{DRL}_{\text{DM}}$ $N_{\text{RX}} =6$};
       		
\addplot+ [mittelgrau, mark options={solid},mark=|,line width=1pt, mark size= 1.5pt] 
				table{%
5000 1.3788810968399048
10000 0.9915938377380371
15000 0.9807305693626404
20000 1.0046686291694642
25000 1.0138962745666504
30000 0.9791227221488953
35000 1.0061727523803712
40000 0.9918033003807067
45000 1.004227912425995
50000 1.0105777025222777
55000 0.981366503238678
60000 1.0074570894241333
65000 1.0084043622016907
70000 0.9938518524169921
75000 0.9998035430908203
80000 0.9734189748764038
85000 1.013385546207428
90000 1.016699993610382
95000 0.97818843126297
100000 1.0133530735969543
105000 1.0237106084823608
110000 0.9831772446632385
115000 1.0123379945755004
120000 0.9948787212371826
125000 0.9957897305488587
130000 0.9797302722930908
135000 1.0029351472854615
140000 1.0172265648841858
145000 0.9872948884963989
150000 1.0118252277374267

			};

            \addlegendentry{\footnotesize $\text{DRL}_{\text{DM}}$ $N_{\text{RX}} =14$};      
            
 \addplot+ [lila, mark options={solid},dashed,line width=1pt,mark=*, mark size= 1.5pt] 
				table{%
5000 14.390387725830077
10000 14.149419403076172
15000 13.776978683471679
20000 14.293248748779297
25000 13.993027877807616
30000 14.213282012939453
35000 14.35253448486328
40000 13.732262802124023
45000 13.822328758239745
50000 14.512628364562989
55000 13.944167327880859
60000 13.800398063659667
65000 14.460885047912598
70000 12.851341247558594
75000 13.95367660522461
80000 13.74853687286377
85000 14.850996398925782
90000 14.067050170898437
95000 13.810523414611817
100000 14.357725715637207

			};

            \addlegendentry{\footnotesize $\text{RNN}_{\text{UNSUP}}$\cite{RNNE2E} $N_{\text{RX}} =14$};  \addplot+ [orange,dashed, mark options={solid},line width=1pt,mark=^, mark size= 1.5pt] 
				table{%
5000 14
10000 14
15000 14
20000 14
25000 14
30000 14
35000 14
40000 14
45000 14
50000 14
55000 14
60000 14
65000 14
70000 14
75000 14
80000 14
85000 14
90000 14
95000 14
100000 14
105000 14
110000 14
115000 14
120000 14
125000 14
130000 14
135000 14
140000 14
145000 14
150000 14
			};
	
            \addlegendentry{\footnotesize $\text{MRC}_{\text{CSI}}$ $N_{\text{RX}} =14$};        	     
    
\addplot+ [rot, solid, mark options={solid},line width=1pt,mark=square*,mark size= 1.5pt] 
				table{%
5000 3.2303431034088135
10000 5.008749485015869
15000 7.87114143371582
20000 8.316414833068848
25000 8.509224891662598
30000 10.630014419555664
35000 11.748169898986816
40000 11.609819412231445
45000 12.76241683959961
50000 12.086512565612793
55000 13.76559829711914
60000 13.128326416015625
65000 14.529611587524414
70000 13.927275657653809
75000 13.91421127319336
80000 14.419867515563965
85000 14.935287475585938
90000 15.277549743652344
95000 15.30213737487793
100000 15.193922996520996
105000 14.971656799316406
110000 15.815604209899902
115000 15.640739440917969
120000 16.70937156677246
125000 17.270212173461914
130000 16.67958641052246
135000 18.355762481689453
140000 16.123849868774414
145000 17.65066146850586
150000 18.69190788269043

			};

            \addlegendentry{\footnotesize $\text{DRL}_{\text{BF}}$ (best) $N_{\text{RX}} = 32$};         
             \addplot+ [mittelblau, solid, mark options={solid},line width=1pt,mark=square*,mark size= 1.5pt] 
				table{
5000 3.152523136138916
10000 5.139096641540528
15000 6.570755052566528
20000 7.53764328956604
25000 8.216304111480714
30000 9.244989967346191
35000 9.273926591873169
40000 9.686792659759522
45000 9.269734954833984
50000 8.789863967895508
55000 9.528003931045532
60000 9.376850461959839
65000 9.37908308506012
70000 9.118344569206238
75000 10.01528160572052
80000 10.169611310958862
85000 9.77604100704193
90000 10.842965340614318
95000 10.411823773384095
100000 10.860155868530274
105000 11.00547480583191
110000 11.702775645256043
115000 11.59183759689331
120000 11.919239926338197
125000 11.96101005077362
130000 12.265054941177368
135000 12.586923146247864
140000 11.682018685340882
145000 12.123906135559082
150000 12.514941215515137

			};

            \addlegendentry{\footnotesize $\text{DRL}_{\text{BF}}$ (avg) $N_{\text{RX}} = 32$}; 
		\end{axis}

	\end{tikzpicture}

%% file: tikz/results/performance_comparison.tikz
\begin{tikzpicture}
		
		\pgfplotsset{compat=1.5}

	\begin{axis}[
	xlabel={SNR [dB]},
	ylabel={Beamforming gain},
	xmajorgrids,
	ymode=log,
	yminorgrids,
	legend style={at={(0.592,0.0),font=\tiny}, anchor=south west}]

			\addplot+ [anthrazit,dashed, mark=o, line width=1.3pt, mark options={ solid}, mark size= 3pt] 
			table{
-15.0 1.639204502105713
-10.0 3.064540147781372
-5.0 5.758774757385254
0.0 15.040102005004883
5.0 23.423490524291992
10.0 30.170106887817383
15.0 31.575841903686523
20.0 32.48244857788086
25.0 32.241477966308594

			};

            \addlegendentry{\footnotesize $\text{RNN}_{\text{UNSUP}}$\cite{RNNE2E}};
	\addplot+ [lila,dashed, mark=square, line width=1.3pt, mark options={ solid}, mark size= 3pt] 
			table{
-15.0 1.479962992668152
-10.0 2.1578585386276243
-5.0 4.385769367218018
0.0 11.327051734924316
5.0 19.648681640625
10.0 27.018993377685547
15.0 30.723038482666016
20.0 31.35284423828125
25.0 32.55327453613281

			};
    
            \addlegendentry{\footnotesize $\text{DNN}_{\text{UNSUP}}$\cite{9367586}};	 	       			    	\addplot+ [apfelgruen,dashed, mark=square*, line width=1.3pt, mark options={ solid}, mark size= 3pt] 
			table{
-15.0 32
-10.0 32
-5.0 32
0.0 32
5.0 32
10.0 32
15.0 32
20.0 32
25.0 32

			};
  
            \addlegendentry{\footnotesize $\text{MRC}_{\text{CSI}}$};	 	   	\addplot+ [orange,dashed, mark=diamond, line width=1.3pt, mark options={ solid}, mark size= 3pt]
			table{
-15.0 1.606712646049567
-10.0 2.589650866661596
-5.0 4.514540926785642
0.0 7.570172340938115
5.0 10.23168086697808
10.0 11.477876664463293
15.0 12.195053349655565
20.0 13.029307910896103
25.0 12.873289201019158

			};
    	
            \addlegendentry{\footnotesize $\text{MRC}_{\text{OMP}}$};	 
              	\addplot+ [mittelgrau,dashed, mark=o*, line width=1.3pt, mark options={ solid}, mark size= 3pt] 
			table{
-15.0 1.467413890361786
-10.0 2.1222286224365234
-5.0 3.126970624923706
0.0 4.13601975440979
5.0 4.510492944717408
10.0 4.433632802963257
15.0 4.515238666534424
20.0 4.551319980621338
25.0 4.493516206741333

			};
    
            \addlegendentry{\footnotesize $\text{Exhaustive}$};	 
            \addplot+ [rot, solid, mark=diamond*, line width=1.3pt, mark options={ solid}, mark size= 3pt] 
			table{

-15.0 1.074355125427246
-10.0 1.2370914220809937
-5.0 3.071650981903076
0.0 5.21556282043457
5.0 9.649805068969727
10.0 13.280217170715332
15.0 17.388103485107422
20.0 19.09268569946289
25.0 19.076297760009766

			};

            \addlegendentry{\footnotesize $\text{DRL}_{\text{BF}}$ (best)};
    \addplot+ [mittelblau, solid, mark=diamond*, line width=1.3pt, mark options={ solid}, mark size= 3pt]
			table{

-15.0 1.1992138385772706
-10.0 1.258617067337036
-5.0 2.2098132848739622
0.0 4.013672947883606
5.0 6.993968486785889
10.0 10.247811865806579
15.0 12.887370610237122
20.0 14.685618710517883
25.0 15.015651965141297

			};
    	
            \addlegendentry{\footnotesize $\text{DRL}_{\text{BF}}$ (avg)};            
		\end{axis}

	\end{tikzpicture}